\documentclass[conference]{IEEEtran}
\usepackage{cite}

\ifCLASSINFOpdf
   \usepackage[pdftex]{graphicx}
   \graphicspath{{../pdf/}{../jpeg/}}
   \DeclareGraphicsExtensions{.pdf,.jpeg,.png}
\else
   \usepackage[dvips]{graphicx}
   \graphicspath{{../eps/}}
   \DeclareGraphicsExtensions{.eps}
\fi
\usepackage[cmex10]{amsmath}
\usepackage{amssymb}
\usepackage{amsfonts}
\usepackage{breqn}

\usepackage{algorithmic}

\usepackage{array}

\usepackage{mdwmath}
\usepackage{mdwtab}

\usepackage{eqparbox}

\usepackage[tight,footnotesize]{subfigure}

\usepackage[caption=false,font=footnotesize]{subfig}
\usepackage{fixltx2e}

\usepackage{stfloats}

\usepackage{url}

\begin{document}
%
\title{Generalized Simplified Variable-Scaled Min Sum LDPC decoder for irregular LDPC Codes}


\vspace{-0.5cm}
\author{\IEEEauthorblockN{Ahmed A. Emran \S \; and Maha Elsabrouty\dag }

\IEEEauthorblockA{
Electronics and Electrical Communications\\
Egypt-Japan University for Science and Technology (E-JUST), Alexandria, Egypt\\
}

Email: \{\S ahmed.emran, \dag maha.elsabrouty\}@ejust.edu.eg}



%


\maketitle

\begin{abstract}
\boldmath
In this paper, we propose a novel low complexity scaling strategy of min-sum decoding algorithm for irregular LDPC codes. In the proposed method, we generalize our previously proposed simplified Variable Scaled Min-Sum (SVS-min-sum) by replacing the sub-optimal starting value and heuristic update for the scaling factor sequence by optimized values. Density evolution and Nelder-Mead optimization are used offline, prior to the decoding, to obtain the optimal starting point and per iteration updating step size for the scaling factor sequence of the proposed scaling strategy. The optimization of these parameters proves to be of noticeable positive impact on the decoding performance. We used different DVB-T2 LDPC codes in our simulation. Simulation results show the superior performance (in both WER and latency) of the proposed algorithm to other Min-Sum based algorithms. In addition to that, generalized SVS-min-sum algorithm has very close performance to LLR-SPA with much lower complexity.
\end{abstract}


%
\IEEEpeerreviewmaketitle

\section{Introduction}
Low-density parity check codes (LDPC) were introduced by Gallager \cite{gallager62low} in the early 1960s. Decoding of LDPC codes, by log-likelihood ratio sum-product algorithms (LLR-SPA), are proven to achieve excellent capacity performance, by approaching the Shannon bound \cite{mackay1999good}. However, the drawbacks of LLR-SPA, namely, the high complexity and sensitivity to linear scaling, are solved by the Min-Sum algorithm \cite{fossorier1999reduced}. 
Scaled Min-Sum \cite{chen2002density} is a modification of Min-Sum algorithm, where a scaling factor is used to decrease the error introduced by using the approximate minimum operation.
Scaled Min-Sum (with constant scaling factor) is suitable for regular LDPC codes. However, irregular LDPC codes require different scaling technique \cite{zhang2005improved,lechner2006improved,ahmed14CCNC}.\\
In \cite{zhang2005improved}, a two-dimensional (2D) correction of the min-sum was proposed. In this algorithm, different scaling factors are required for different check node degrees and variable node degrees. Consequently, the algorithm requires the calculation of two scaling factor vectors \underline{\textbf{$\alpha$}} and  \underline{\textbf{$\beta$}} with length equal maximum check node's degree and variable node's degree respectively. These vectors are optimized by parallel differential optimization of Density Evolution (DE).\\
In \cite{lechner2006improved}, different scaling factor per iteration is proposed for irregular LDPC codes. Different scaling factor per iteration technique has good performance for irregular LDPC codes. However, adding these scaling factors requires complex calculation steps in designing stage, and requires extra storage to store the scaling factor value of each iteration.\\
In \cite{ahmed14CCNC}, we proposed simplified variable-scaled min-sum (SVS-min-sum) decoding technique. This algorithm uses simply implemented heuristic technique to update the scaling factor with iterations. It is simpler than both variable scaling factor \cite{lechner2006improved} and 2D correction Min-Sum \cite{zhang2005improved} in implementing and designing. Simulation results show that SVS-min-sum has lower Bit Error Rate (BER) than constant scaling factor for many LDPC codes \cite{ahmed14CCNC}. SVS-min-sum algorithm starts the scaling factor sequence with a constant value equals 0.5. This restriction decreases its performance and makes it unsuitable for some codes.\\
In this paper, we introduce a generalization of the SVS-min-sum algorithm by removing the restriction of starting the scaling sequence from 0.5. This generalization leads to better performance than constant scaling for all codes. In fact, constant scaling can be seen as a special sub-optimized version of the proposed algorithm as shown in section IV. 
 We apply Nelder-Mead optimization \cite{nelder1965simplex} on DE to jointly optimize the initial scaling factor and updating step of the scaling sequence. Simulation results illustrate the improvement of the proposed algorithm in both BER performance and decoding latency over other scaling strategies.\\
The rest of the paper is organized as follows: Section II presents the necessary background on the SPA, Min-Sum, Scaled Min-Sum, Variable Scaled Min-Sum and SVS-min-sum algorithms. Section III presents the generalized SVS-min-sum algorithm. The simulation results are displayed and discussed in Section IV. Finally, the paper is concluded in section V.\\

\section{REVIEW OF THE SPA AND MIN-SUM based ALGORITHMS}
An $(n,k)$  LDPC code is a binary code characterized by a sparse parity check matrix $\mathbf{H} \in \mathbb{F}_2^{m\times n}$ where $m=n-k$. It can be represented by a Tanner graph which contains variable nodes $j \in \{1..n\}$ and check nodes $i\in \{1..m\} $ . We denote the set of variable nodes connected to a certain check node $i$ as $V\{i\}$. Furthermore, the set $V\{i\}/j$ denotes the set of variable nodes connected to check node $i$ excluding $j$. Similarly, the set of check nodes connected to a certain variable node $j$ is denoted by $C\{j\}$. $C\{j\}/i$ denotes the set of check nodes connected to the variable node $j$ excluding $i$. \\
LDPC codes are efficiently decoded by message passing decoding algorithms. The main idea behind all message passing algorithms is processing the received symbols iteratively in concatenated steps that can be seen over the Tanner graph as horizontal step followed by vertical step. In this section, we review some message passing decoding algorithms that are either used for comparision or as the starting point of our modified algorithm.

\subsection{Sum-Product algorithm (SPA)}
One iteration of the tanh-based SPA is described in the following steps:-
\begin{enumerate}
  \item \textit{Initialization step:}
The $LLR$ of bit number $j$ is initialized with its channel $LLR$ $(U_{ch_j})$. These initial values are used as $v_{j \to i}$ , messages from variable node $j$ to check nodes $i$ $\forall$ $i \in C\{j\}$.
  \item \textit{Horizontal step: }
  At each check node $i$, messages $v_{j \to i}$ (which come from variable nodes $V\{i\}$) are used to calculate the reply messages $U_{i \to j}$  for all $j \in V\{i\}$ by \eqref{eq:1}.
\begin{equation} \label{eq:1}
U_{i \to j}=2 \times tanh^{-1} \bigg( \prod_{\substack{j' \in V\{i\}/j}} tanh\frac{v_{j' \to i}}{2}\bigg)
\end{equation}
  \item \textit{Vertical step: }
At each variable node $j$, messages $U_{i \to j}$ are used to calculate the reply messages $v_{j \to i}$  for all $i \in C\{j\}$ by \eqref{eq:2}.
  	\begin{dmath} \label{eq:2}
  		v_{j \to i}=U_{ch_j}+\sum_{\substack{i'\in C\{j\}/i}} U_{i'\to j}
	\end{dmath}  
  
  \item Decision step\\
For each variable node $j$, its $LLR$ is updated by \eqref{eq:3}
  	\begin{dmath} \label{eq:3}
  		LLR_{j}=U_{ch_j}+\sum_{\substack{i'\in C\{j\}}} U_{i'\to j}
	\end{dmath}  
\end{enumerate}
The LLR values are applied to the hard decision to decide on the bit value to be 1 if $LLR_{j}<0$  and zero otherwise. The syndrome is calculated and checked; if it is all-zero vector, this word is successfully decoded, otherwise, if the syndrome condition is not satisfied, the decoder proceeds to the next iteration. This process continues till either the code word is successfully decoded or the maximum iterations are exhausted. 

\subsection{Min-Sum algorithm}
The Min-Sum algorithm follows the same steps as SPA. It only approximates the horizontal step calculation by minimum operation as shown in \eqref{eq:4} \cite{fossorier1999reduced}
  	\begin{dmath} \label{eq:4}
  		U_{i \to j}=\prod_{\substack{j'\in V\{i\}/j}} sign(v_{j'\to i})*\min_{\substack{j' \in V\{i\}/j}} {|v_{j' \to i}|}
	\end{dmath}  
Min-Sum is easier to implement, as it gets rid of the tanh(.) calculation. However, the approximation of the tanh(.) to the min(.) leads to some loss of performance compared to the tanh-based SPA algorithm. This loss of performance is partially recovered by Scaled Min-Sum algorithm.

\subsection{Scaled Min-Sum algorithm}
In order to decrease the gap between the min-sum and the tanh-based SPA algorithms, a constant scaling factor ($\alpha < 1$) is applied to the check node updating equation \eqref{eq:4}. In other words, converts the Horizontal step to \eqref{eq:5}
  	\begin{dmath} \label{eq:5}
  		U_{i \to j}=\alpha \prod_{\substack{j'\in V\{i\}/j}} sign(v_{j'\to i})*\min_{\substack{j' \in V\{i\}/j}} {|v_{j' \to i}|}
	\end{dmath} 
This scaling factor is optimized to maximize the performance of Scaled Min-Sum algorithm.

\begin{figure}[!t]
\centering
\includegraphics[width=3.5in]{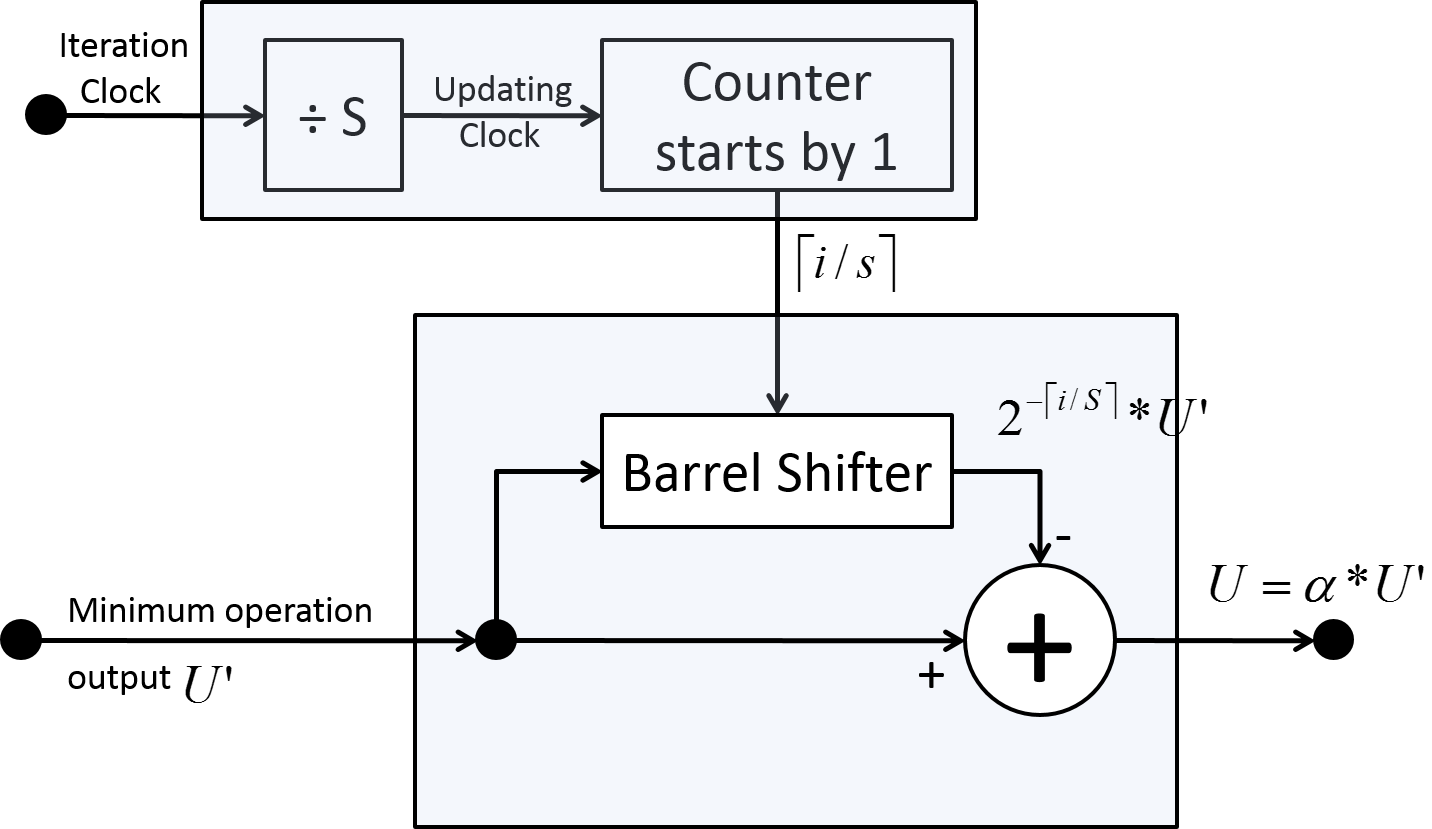}
\caption{Circuit representation of SVS scaling}
\label{SVS_block}
\end{figure}

\begin{table}[!t]
\renewcommand{\arraystretch}{1.3}
\caption{Calculation of scaling factor value of each iteration using SVS-min-sum algorithm}
\label{table_I}
\centering
\begin{tabular}{|c|c|}
\hline
Iteration index $i$ & $\alpha$ \\
\hline
$1 \to S$ & 0.5 \\
\hline
$(S+1) \to 2S$ & 0.75 \\
\hline
$(2S+1) \to 3S$ & 0.875\\
\hline
$(3S+1) \to 4S$ & 0.9375 \\
\hline
\end{tabular}
\end{table}

\subsection{SVS-min-sum algorithm} \label{sub:SVS}
Changing the scaling factor with iteration for irregular LDPC codes is used in \cite{lechner2006improved}. Despite of performance enhancement of this variable scaled min-sum algorithm, it requires extra storage because we need different scaling factor value per iteration, associated with different mutual information into passed messages per iteration \cite{lechner2006improved}. The general fractional values (taken by the scaling factors) make the multiplication operation complex to implement. We proposed in \cite{ahmed14CCNC} an SVS-min-sum algorithm addresses the particular point of simply per-iteration updated scaling rule.\\ 
As stated in \cite{lechner2006improved}, the scaling factor should increase exponentially with iterations and its final value is 1. So we approximate the scaling factor sequence to a stair sequence which is updated every $S$ iterations, increase exponentially and easy to implement. The variable scaling factor can be calculated as:  
	\begin{equation}	 \label{eq:6}
		\alpha=1-2^{-\lceil i/S \rceil}
	\end{equation}
Where $\lceil i/S \rceil$ is the first integer greater than or equal to $i/S$ . $i$ is the iteration index which takes values $\{1, 2, 3 ...\}$. By using \eqref{eq:6}, the scaling factor of each iteration can be calculated as shown in table \ref{table_I}. This sequence is:-
\begin{enumerate}
  \item Easy to design, because it requires a single parameter $S$.
  \item Does not need to store a specific scaling sequence for each code rate.  It only requires to store the optimal updating step size $S$ of each code rate.
  \item Easy to implement, because it only requires shifting right by $\lceil i/S \rceil $ then subtraction. Number of required shifts $\lceil i/S \rceil $ can be stored in a register and increased by 1 every $S$ iterations.
\end{enumerate}
As shown in Fig. \ref{SVS_block}, the SVS scaling strategy is implemented by two sub-circuits. The first sub-circuit calculates the number of shifts required in each iteration $\lceil i/S \rceil $, this number of shifts is calculated by dividing the iteration clock by $S$ (updating step) to generate the updating clock of the scaling factor, then this updating clock is used to count the required number of shifts $\lceil i/S \rceil $ using a counter starts by 1. The other sub-circuit multiplies the minimum operation output $U'$ by $\alpha$ specified in \eqref{eq:6} by using a Barrel shifter to shift $U'$ right by $\lceil i/S \rceil $ then subtract.

\begin{figure}[!t]
\centering
\includegraphics[width=3.5in]{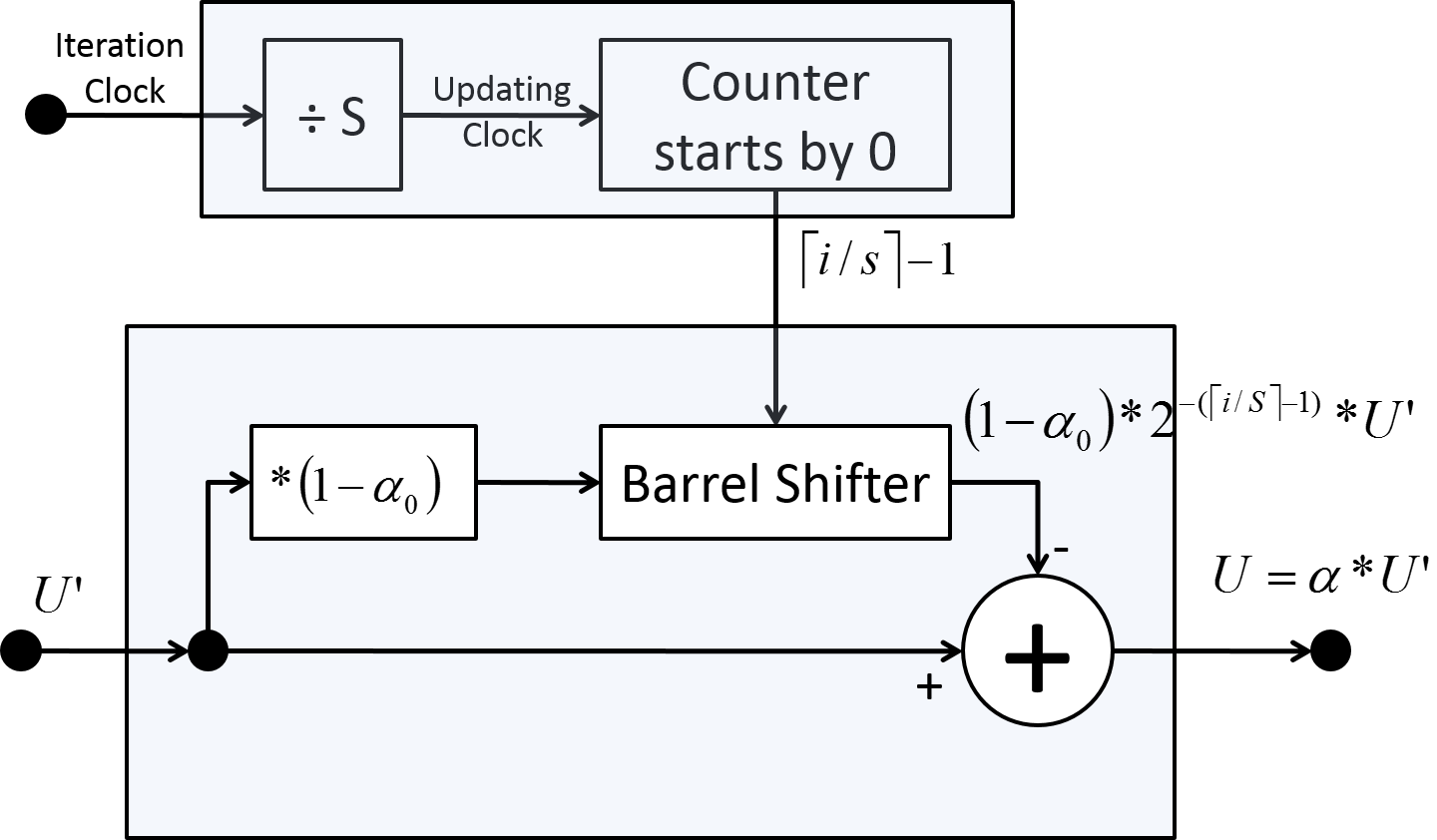}
\caption{Circuit representation of GSVS scaling}
\label{GSVS_block}
\end{figure}
\begin{table}[!t]
\renewcommand{\arraystretch}{1.3}
\caption{Calculation of scaling factor value of each iteration using GSVS-min-sum algorithm}
\label{table_II}
\centering

\begin{tabular}{|c|c|}
\hline
Iteration index $i$ & $\alpha$ \\
\hline
$1 \to S$       &  $\alpha_{0}$ \\
\hline
$(S+1) \to 2S$  &  $0.5+ 0.5*\alpha_{0}$ \\
\hline
$(2S+1) \to 3S$ &  $0.75+ 0.25*\alpha_{0}$\\
\hline
$(3S+1) \to 4S$ &  $0.875+ 0.125*\alpha_{0}$ \\
\hline
\end{tabular}
\end{table}

\section{Generalized SVS-min-sum (GSVS-min-sum) algorithm}
As shown in table \ref{table_I}, SVS-min-sum sequence starts with 0.5 and increases exponentially with iterations. The limitation of starting with a fixed value of 0.5 restricts the performance to be sub-optimal. As a solution, we propose a new GSVS-min-sum algorithm where the scaling factors sequence is calculated by \eqref{eq:7}:
	\begin{equation}  \label{eq:7}
		\alpha=1-(1-\alpha_{0})*2^{-(\lceil i/S \rceil-1)}
	\end{equation}
Where  $\alpha_{0}$ is the initial scaling factor. By using \eqref{eq:7}, scaling factor of each iteration is calculated as shown in table \ref{table_II}, where scaling factor values start with $\alpha_0$ and increase exponentially to unity for large value of iteration index $i$.\\
Circuit representation of GSVS scaling is shown in Fig. \ref{GSVS_block}. It is similar to SVS scaling circuit, but with two main differences: the first difference is that $U'$ is multiplied by $(1-\alpha_0)$ before shifting right, this is added to generalize the initial scaling factor.  $(1-\alpha_0)$ is chosen to be simply implemented. We use $(1-\alpha_{0})$ to be in the form of $2^{-i}$ or $2^{-j}+2^{-k}$, where $i$, $j$ and $k$ are integer numbers; for example, if ${i=2 \to (1-\alpha_{0})=0.25}$, ${i=3 \to (1-\alpha_{0})=0.125}$ and if ${j=2,k=3 \to (1-\alpha_{0})=0.375}$ . The second difference is that the counter of required shifts starts with 0 instead of 1 because GSVS-min-sum requires $(\lceil i/S \rceil-1)$ shifts not $\lceil i/S \rceil$ as in SVS-min-sum.\\
In SVS-min-sum, we only need to optimize the updating step size $S$, however, in GSVS-min-sum we also need to optimize the initial scaling factor $\alpha_0$. To calculate the optimal $(\alpha_{0}, S)_{opt.}$, we use Nelder-Mead optimization Method \cite{nelder1965simplex} (summarized in \ref{sub:NM}) to minimize $(E_b/N_0)_{min}$, where $(E_b/N_0)_{min}$ is the minimum $E_b/N_0$ required to achieve pre-specified BER threshold. $(E_b/N_0)_{min}$ of given $(\alpha_{0}, S)$ is calculated by DE.

\subsection{Density Evolution (DE) of Min-Sum based algorithms} \label{sub:DE}
DE checks the ability of an LDPC decoder to correctly decode messages with specific noise variance. This is done by tracing the Probability Density Function (PDF) of messages passed between check and variable nodes (using all-zero code-word). Using of all-zero code-word is valid in Binary Phase Shift Keying (BPSK) because the LLR of both 1 and 0 has a similar PDF shape, but this is not valid in Quadrature Amplitude Modulation (QAM) signaling \cite{tullberg2005serial}.\\
DE is used in \cite{mackay1999good} to obtain the optimal weight distribution of irregular LDPC codes, and is used in \cite{chen2002density} \cite{zhang2005improved} to calculate the optimal scaling factor(s) of LDPC decoder. We use the same DE as in \cite{zhang2005improved} after changing the PDF of channel LLR so that we can use it with QAM signaling. 

\subsubsection{Channel LLR's PDF of BPSK over an AWGN channel}
In BPSK, all-zero code-word's bits $V_{k}=0$ are modulated to $x_{k}=1-2V_{k}=1$. Then $x_{k}$ is transmitted over an Additive White Gaussian Noise (AWGN) channel with noise variance $\sigma^{2} $, so the received sequence is $y_{k}=1+n_{k}$ where $n_{k}$ is normally distributed random variable with mean=0 and variance=$\sigma^{2} $ . Therefore, the channel LLR $(U_{ch_{k}}=2 \times y_{k}/\sigma^{2})$ is normally distributed random variable with mean=$2/\sigma^{2}$ and variance=$4/\sigma^{2}$. The PDF of $U_{ch_{k}}$ is used as the initial PDF of variable nodes' messages to check nodes. 

\subsubsection{Channel LLR's PDF of higher order QAM constellation over an AWGN channel}
As shown in \cite{tullberg2005serial}, for higher order constellations, we cannot assume that all-zero code-word was transmitted. Therefore, we used a similar procedure to \cite{tullberg2005serial}, where authors modified the definition of bit's LLR to be: "\textit{LLR of receiving the same bit value as was transmitted}" instead of "\textit{LLR of receiving 0}". This is equivalent to replacing $U_{ch}$ by $U_{ch}^{+}$,  where  $U^{+}_{ch_k}=U_{ch_k}\times(1-2V_{k})$. So $U^{+}_{ch_k}$ will be positive if and only if $U_{ch_k}$ has the same sign as given by $V_{k}$ .\\
Firstly, for any constellation point $W$, we calculate the PDF of $U_{ch}$ for bit number $l$ into $W$ given that $W$ is transmitted $f_{U_{ch}}(u_{l}/W)$ \cite{benjillali2006probability}. Then we calculate the average PDF of $U_{ch}^{+}$ by \eqref{eq:8}. 
	\begin{align} \label{eq:8}
	f_{U_{ch}^{+}}(u)=\frac{1}{\eta} \sum_{l=1}^{log_{2}(M)/2} \bigg( &\sum_{W \in Z_{l}} f_{U_{ch}}(u_{l}=u/W)\nonumber \\ + & \sum_{W \in O_{l}} f_{U_{ch}}(u_{l}=-u/W) \bigg)
	\end{align}
Where $l$ is the bit position index. Only half of bit positions were used, because of symmetry between real and imaginary axes of QAM signaling. $Z_{l}$ is the set of $W$ where bit position $l$ contains 0. $O_{l}$ is the set of $W$ where bit position $l$ contains 1. $\eta$ is PDF correction factor used to ensure that area under the PDF=1.\\
In other words, calculate the average PDF of $U_{ch}$ for zeros and $-U_{ch}$ for ones. Then use this PDF as the initial PDF of variable nodes' messages.

\subsection{Nelder-Mead (NM) optimization method } \label{sub:NM}
\begin{figure}[!t]
\centering
\includegraphics[width=3.5in]{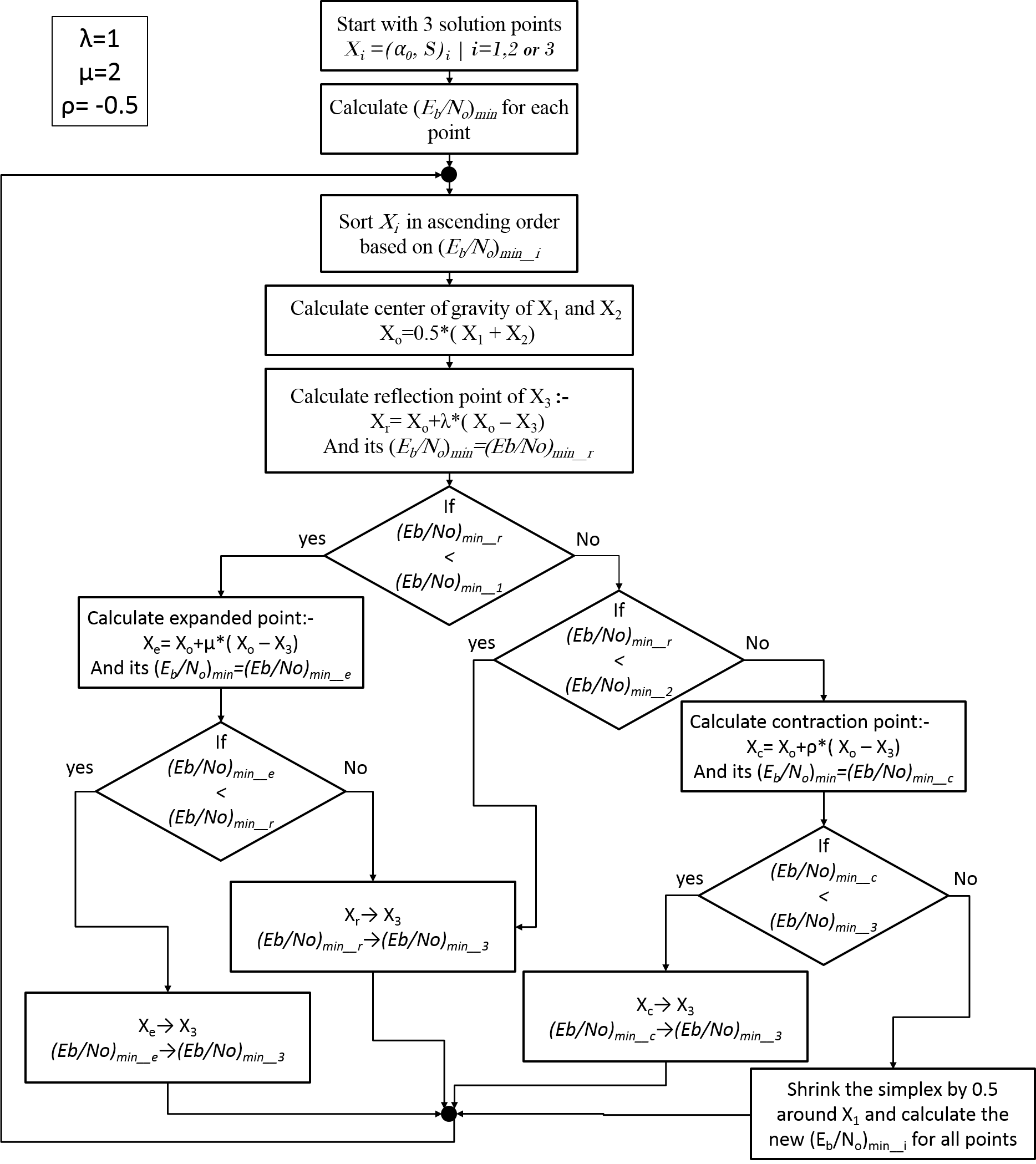}
\caption{Flow chart of Nelder-Mead method}
\label{NM_block}
\end{figure}

\begin{figure}[!t]
\centering
\includegraphics[width=2.5in]{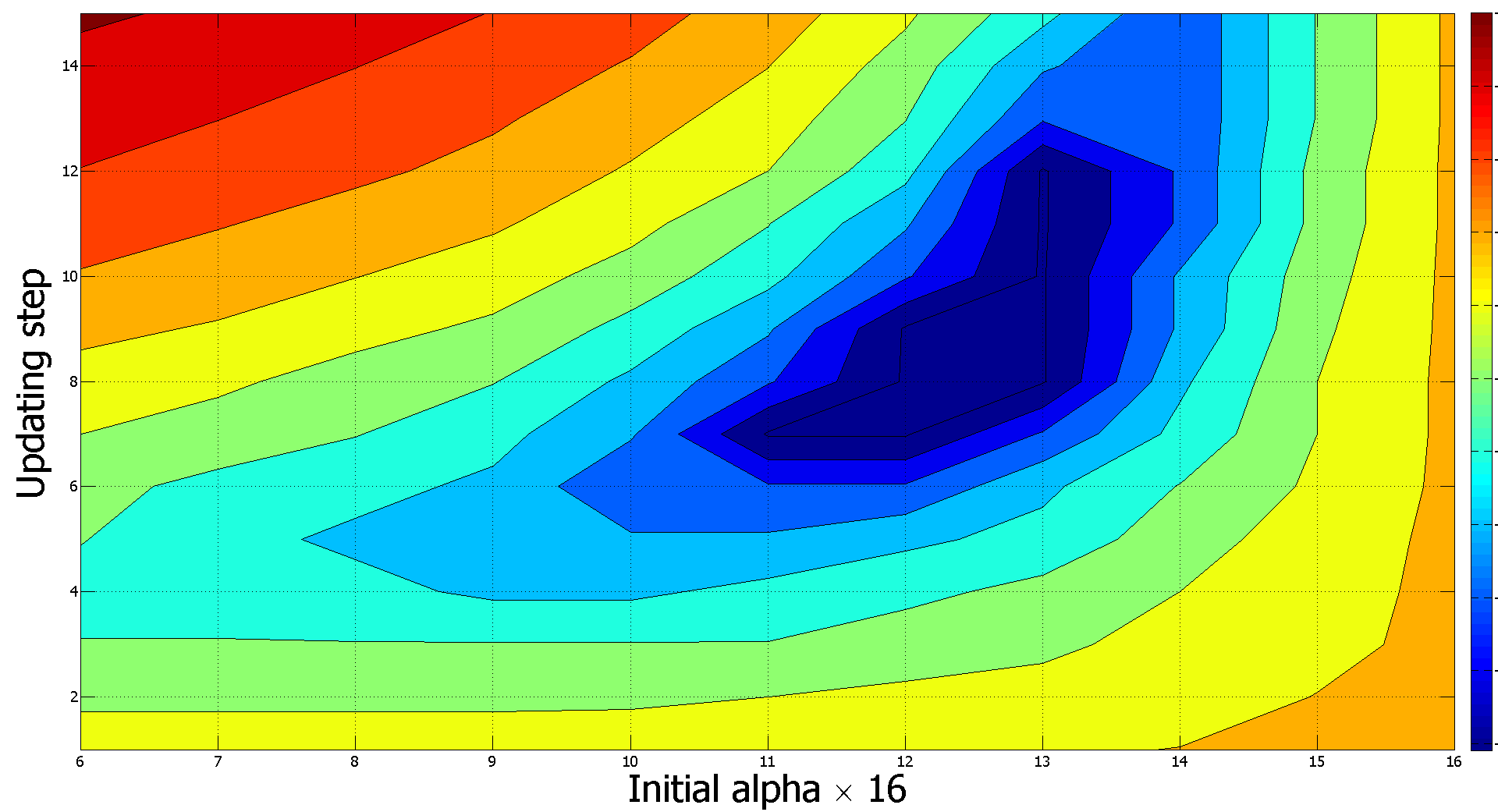}
\caption{contours of $(E_b/N_0)_{min}$ for short code with rate=0.5}
\label{single_min}
\end{figure}

NM optimization \cite{nelder1965simplex} of $(\alpha_{0}, S)$ is based on constructing a simplex (polygon) of 2+1=3 random solution points $\{X_i =(\alpha_0 , S)_i | i=1,2$ or $ 3\}$ for our 2-dimension problem. After calculating $(E_b/N_0)_{min}$ of these three points, the worst point is replaced by a better point as described in the flow chart in Fig.\ref{NM_block}. This procedure is repeated until the simplex shrink enough to the optimal solution.\\
We use Nelder-Mead method for many reasons: first, it does not need the mathematical derivative of the cost function (which is not available because our cost function is $(E_b/N_0)_{min}$ which comes from DE). Second, Nelder-Mead is faster in convergence than other heuristic methods. Finally, our cost function has only one minimum, so the algorithm does not get trapped in a local minimum. To prove the last claim that $(E_b/N_0)_{min}$ has only one minimum, we calculated it for all possible combinations of $(\alpha_0, S)$ for short length LDPC code with rate 0.5 specified in DVB-T2. Fig.\ref{single_min} shows that $(E_b/N_0)_{min}$ has only one minimum. Similar results are obtained for all tested codes. Optimization of $(\alpha_0, S)$ is calculated offline for each LDPC code rate, then the optimal value is used to implement the decoding circuit.
\section{Simulation environment and results}
For simulation, we used (16200, 7200) eIRA LDPC code specified in DVB-T2 standard \cite{bluebooka122, bluebooka133}. Data are produced as binary bits modulated using the challenging 256-QAM modulation scheme and sent over an AWGN channel. The simulations are performed using MATLAB platform. Maximum number of iterations is set to 40 iterations.

Fig.\ref{WER_05_Short} shows the WER of LLR-SPA, SVS-min-sum with $S=10$ \cite{ahmed14CCNC}, Scaled Min-Sum with $\alpha= 15/16$ (optimized by DE and the same as in \cite{ahmed14CCNC}), GSVS-min-sum with $(\alpha_0=0.75$ and $S=9)$ (optimized by DE with Nelder-Mead method) and 2D correction min-sum; where the output of the check nodes with degree 4,5,6 and 7 is multiplied by 0.94, 0.92, 0.88 and 0.86 respectively, and the output of the variable nodes with degree 1,2,3 and 8 is multiplied by 1.00, 1.00, 0.91 and 0.83 respectively \cite{zhang2005improved}. Results in Fig.\ref{WER_05_Short} show that: 

\begin{figure}[!t]
\centering
\includegraphics[width=3.5in]{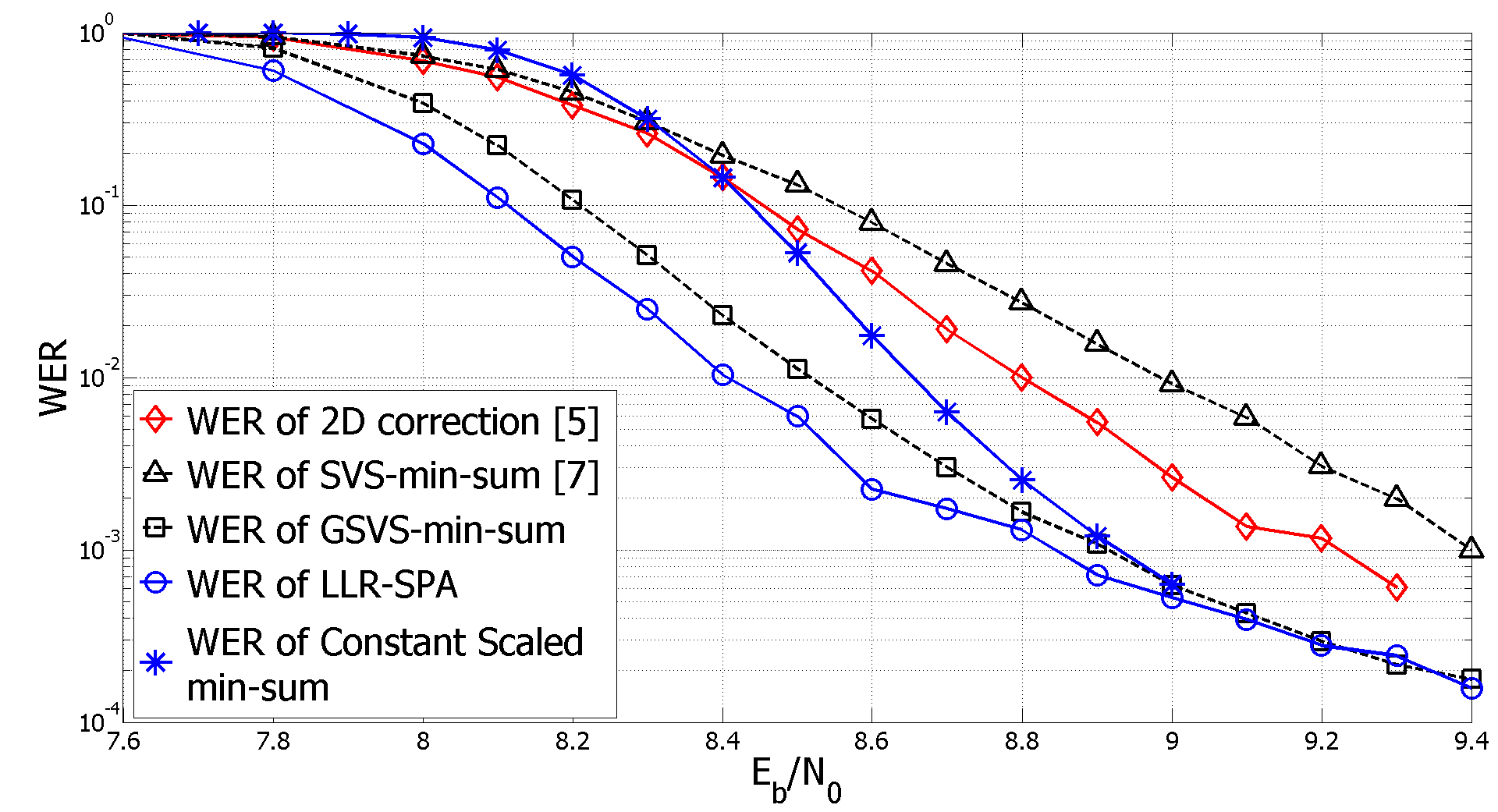}
\caption{WER of (16200, 7200) LDPC over 256QAM with different decoding algorithms}
\label{WER_05_Short}
\end{figure}

\begin{itemize}
\item GSVS-min-sum has better performance than SVS-min-sum by 0.5 dB at WER= $10^{-3}$, this indicates the importance of the proposed algorithm, which jointly optimizes the initial scaling factor $\alpha_0$ with the updating step size $S$. 
\item Although 2D correction min-sum has an excellent performance after many iterations (200 iterations) \cite{zhang2005improved}, it has higher WER than GSVS-min-sum algorithm after 40 iterations for the whole simulation range and higher than scaled-min-sum for high $E_b/N_0$ range. The gap between GSVS-min-sum and 2D correction is 0.3 dB at WER= $10^{-3}$.
\item There is a small gap between GSVS-min-sum and LLR-SPA performances (0.1 dB for low $E_b/N_0$ and nearly disappears at high $E_b/N_0$), with much lower implementation complexity.
\item For the scaled min-sum algorithm, Optimizing the scaling factor of each code rate increases its performance specially for high $E_b/N_0$. This concept is illustrated in \cite{ahmed14CCNC} by showing that each code rate of DVB-T2 LDPC codes has different optimal scaling factor.
\end{itemize}

\begin{figure}[!t]
\centering
\includegraphics[width=3.5 in]{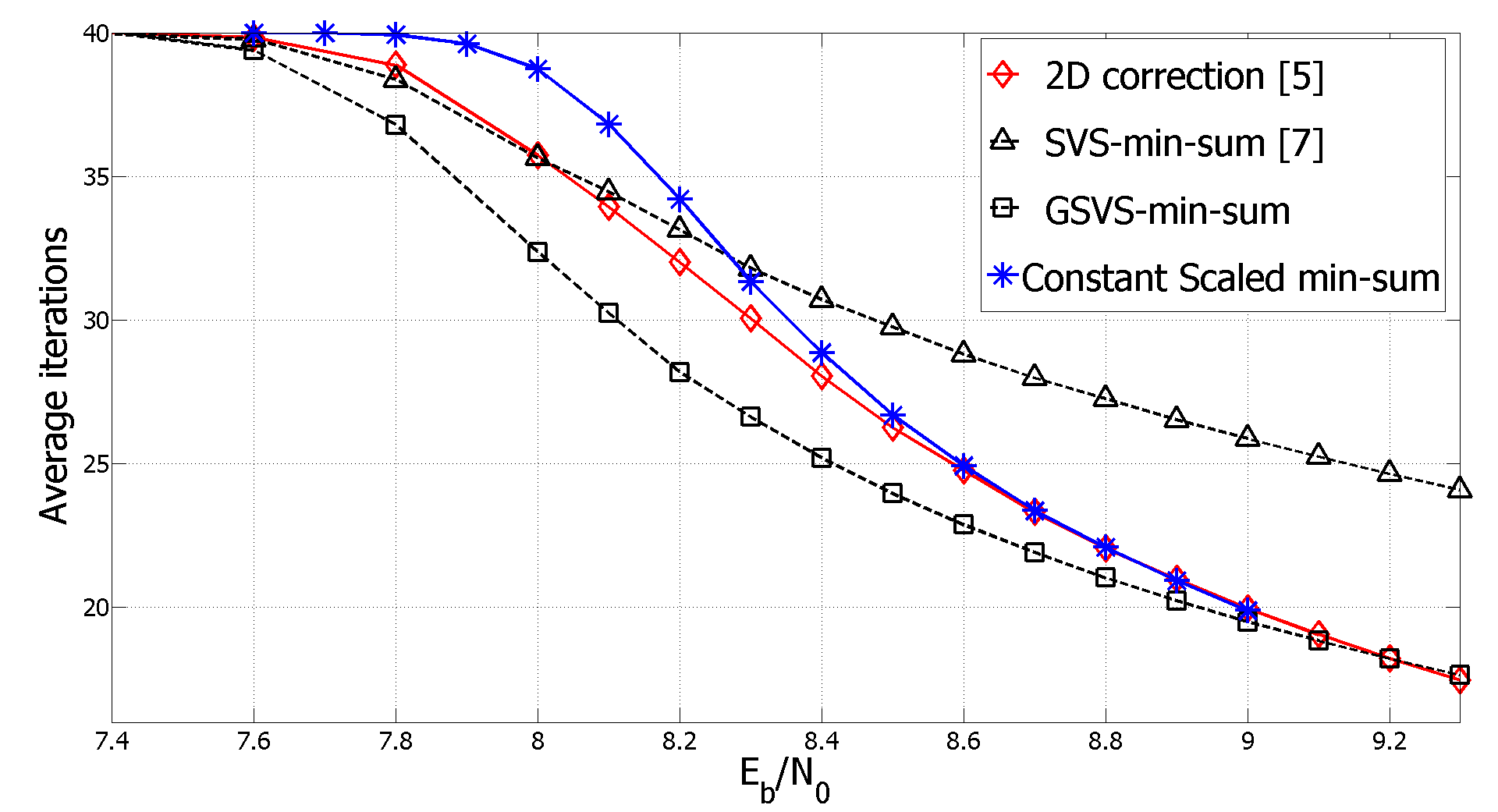}
\caption{Average number of iterations of (16200, 7200) LDPC over 256QAM with different decoding algorithms}
\label{Itr_05_Short}
\end{figure}

\begin{figure}[!t]
\centering
\includegraphics[width=3.5in]{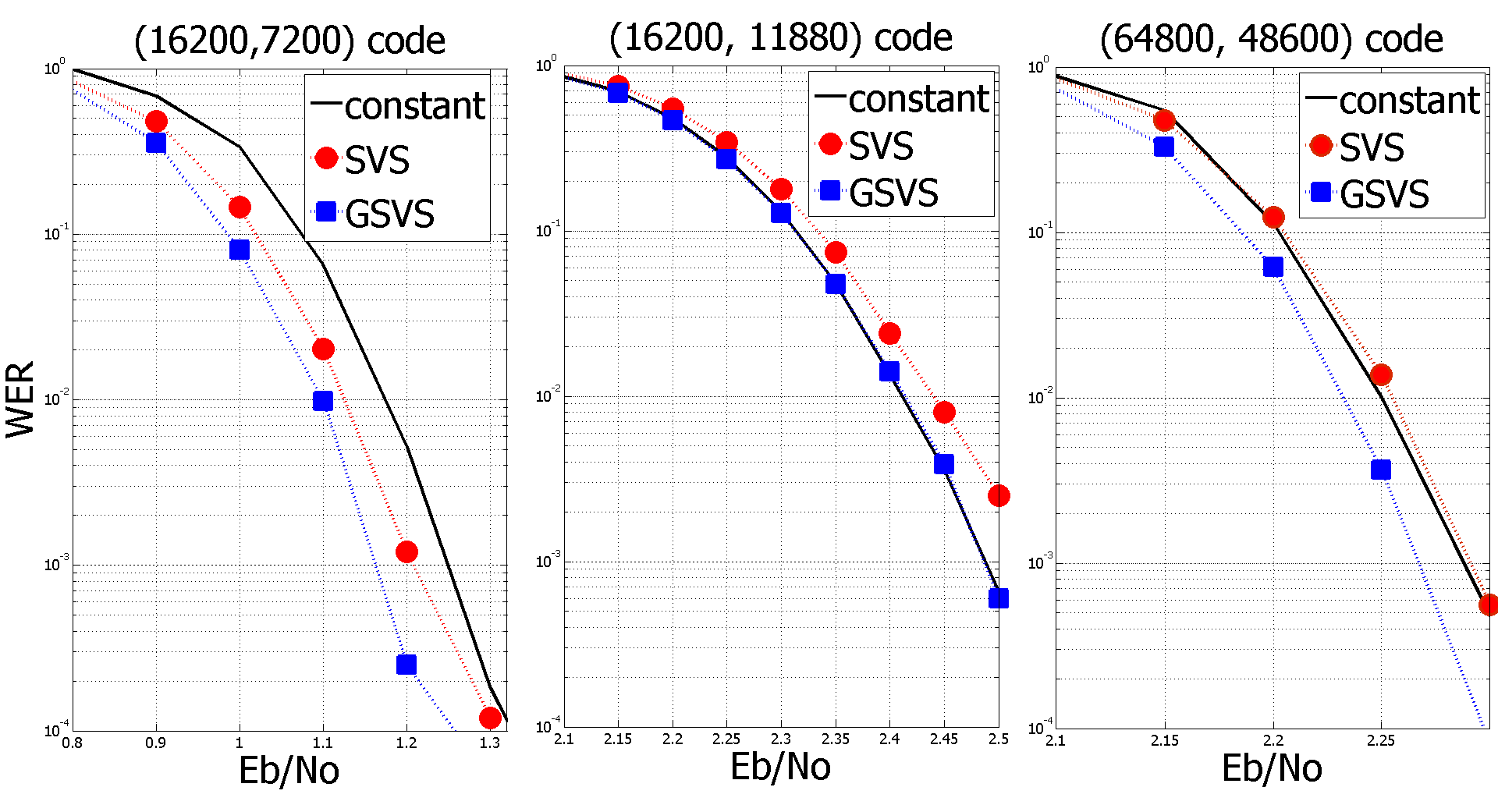}
\caption{WER of different LDPC cods over BPSK}
\label{WER_BPSK}
\end{figure}

Fig.\ref{Itr_05_Short} shows clearly that GSVS-min-sum not only has lower WER than other min-sum based algorithms, but also it has the lowest average number of iterations which leads to lower latency and higher average throughput.
\\For more results, we used three different rates of the LDPC codes specified in DVB-T2 standard with BPSK, these codes are (16200,7200) short code with nominal rate = 0.5, (16200,11880) short code with nominal rate = 0.75 and (64800,48600) normal code with rate = 0.75. WER of these codes with constant scaled-min-sum, SVS-min-sum and GSVS-min-sum decoding algorithm are shown in Fig. \ref{WER_BPSK}. Simulation results show that GSVS-min-sum has the lowest WER among the three decoding algorithms, even though scaled min-sum has lower WER than SVS-min-sum or not. Note that: constant scaling (in scaled min-sum) and SVS-min-sum are special cases of the GSVS-min-sum where $S$= number of iterations for scaled min-sum and $\alpha_0=0.5$ for SVS-min-sum. So GSVS-min-sum, which has optimized values for both $\alpha_0$ and $S$, has the best performance between them as shown in Fig. \ref{WER_BPSK}. The parameters of the three decoding algorithms are shown in table \ref{table_III}. For (16200,11880) code, GSVS-min-sum has the same performance as constant scaling factor and better performance than SVS-min-sum. The poor performance of SVS-min-sum comes from the limitation of starting by 0.5 which is away from the optimal scaling sequence. For the other codes, GSVS-min-sum has better performance than both SVS-min-sum and constant scaled min-sum.

\section{CONCLUSION AND FUTURE WORK}
In this paper, we generalized the SVS-min-sum decoder by allowing it to start with any initial scaling factor $\alpha_0$. Simulation results indicated the superior performance  and lower latency of GSVS-min-sum decoder to other min-sum based algorithms. In addition, GSVS-min-sum algorithm performance is very close to the LLR-SPA with much lower complexity.  Moreover,  the proposed algorthim is still simpler to implement than both the variable scaling factor in \cite{lechner2006improved} and the 2D correction Min-Sum in \cite{zhang2005improved}. As future work, we will apply our GSVS-min-sum decoding algorithm to layered LDPC codes implementation.

\section*{Acknowledgment}
This work is supported by E-JUST and Minister of High Education (MoHE). And is supported by NTRA as part of the project "Design and implementation of DVB-T/T2 solution". 

\begin{table}[!t]
\renewcommand{\arraystretch}{1.3}
\caption{Optimal parameters of different LDPC decoding algorithms for BPSK}
\label{table_III}
\centering
\begin{tabular}{|c|c|c|c|}
\hline
  & constant $\alpha$ & step of SVS & $(\alpha_0,S)$ of GSVS \\
\hline
$(16200,7200)$ & 0.9375 & 5 & $(0.75,9)$\\
\hline
$(16200,11880)$ & 0.875 & 10 & $(0.75,16)$ \\
\hline
$(64800,48600)$ & 0.875 & 10 & $(0.75,18)$\\
\hline
\end{tabular}
\end{table}

\bibliographystyle{IEEEtran}
\bibliography{papers}
\end{document}